\documentclass[showpacs,showkeys,11pt,
preprint,preprintnumbers,nofootinbib,
groupedaddress,superscriptaddress]{revtex4}

\usepackage{amsmath,amssymb}	% required for `\align' (yatex added)
\usepackage[dvipdfm]{graphicx}
\usepackage{cancel}
\usepackage[hypertex]{hyperref}

\begin{document}
%%%%%%%%%%%%%%%%%%%%%%%%%%%%%%%%%%%%%%%%%%%%%%%%%%%%%%%%%%%
\title{Determination of $\tan\beta$ from the Higgs boson decay at linear
colliders} 
\preprint{UT-HET-079}
%\keywords{Higgs boson}
%%%%%%%%%%%%%%%%%%%%%%%%%%%%%%%%%%%%%%%%%%%%%%%%%%%%%%%%%%%
\author{Shinya Kanemura}
\email{kanemu@sci.u-toyama.ac.jp}
\affiliation{Department of Physics, University of Toyama, Toyama
930-8555, Japan} 
%%%%%%%%%%%%%%%%%%%%%%%%%%%%%%%%%%%%%%%%%%%%%%%%%%%%%%%%%%%
\author{Koji Tsumura}
\email{ko2@eken.phys.nagoya-u.ac.jp}
\affiliation{Department of Physics, Nagoya University, Nagoya 464-8602,
Japan} 
%%%%%%%%%%%%%%%%%%%%%%%%%%%%%%%%%%%%%%%%%%%%%%%%%%%%%%%%%%%
\author{Hiroshi Yokoya}
\email{hyokoya@sci.u-toyama.ac.jp}
\affiliation{Department of Physics, University of Toyama, Toyama
930-8555, Japan} 
%%%%%%%%%%%%%%%%%%%%%%%%%%%%%%%%%%%%%%%%%%%%%%%%%%%%%%%%%%%

%\date{\today}

%%%%%%%%%%%%%%%%%%%%%%%%%%%%%%%%%%%%%%%%%%%%%%%%%%%%%%%%%%%
\begin{abstract}
In the two Higgs doublet model, $\tan\beta$ is an important parameter,
 which is defined as the ratio of the vacuum expectation values of the
 doublets. 
We study how accurately $\tan\beta$ can be determined at linear
 colliders via the precision measurement of the decay branching fraction of
 the standard model (SM) like Higgs boson.
Since the effective coupling constants of the Higgs boson with the weak
 gauge bosons are expected to be measured accurately, the branching
 ratios can be precisely determined.
Consequently, $\tan\beta$ can be determined with a certain amount of
 accuracy.
Comparing the method to those using direct production of the additional
 Higgs bosons, we find that, depending on the type of Yukawa
 interactions, the precision measurement of the decay of the SM-like
 Higgs boson can be the best way to determine $\tan\beta$, when the
 deviations in the coupling constants with the gauge boson from the SM
 prediction are observed at linear colliders.
\end{abstract}
%%%%%%%%%%%%%%%%%%%%%%%%%%%%%%%%%%%%%%%%%%%%%%%%%%%%%%%%%%%
\maketitle

%%%%%%%%%%%%%%%%%%%%%%%%%%%%%%%%%%%%%%%%%%%%%%%%%%%%%%%%%%%
\section{Introduction}

% Introduction
A Higgs boson has been discovered at the LHC~\cite{Ref:atlas,Ref:cms}.
Current data show that its properties such as its mass, production cross
sections times the decay branching ratios and the spin-parity are consistent
with those of the Higgs boson in the standard model
(SM)~\cite{Ref:atlas-comb,Ref:cms-comb}. 
However, the whole structure of the Higgs sector has not been clarified
at all. 
Since there is no principle to determine the structure of the Higgs
sector, the SM Higgs sector is the simplest but just one of the
possibilities.
There are many problems, which should be explained by new
physics beyond the SM, such as the naturalness problem, the origin of
tiny neutrino masses and mixings, the existence of dark matter, etc. 
Various extensions of the SM considered to solve these problems
often contain the extended Higgs sector, where new Higgs multiplets are
added to the SM Higgs sector. 

% THDM and new physics
Multi Higgs models are constrained by the electroweak $\rho$ parameter 
significantly.
The two Higgs doublet model (THDM) is the natural and minimal extension
of the SM Higgs sector, since multi Higgs doublet models predict the
$\rho$ parameter to be unity at the tree level~\cite{Ref:HHG}.
In general, the THDM predicts the flavor changing neutral current
(FCNC) which is severely constrained by the experimental data.  
This problem may be solved by introducing a discrete $Z_2$ symmetry under which 
the different parity is assigned to each doublet field~\cite{Ref:GW}.
Under this symmetry, each fermion couples with only one Higgs doublet, 
and hence the FCNC is absent at the tree level.
Depending on the assignment of the $Z_2$ parity to each fermion, there
are four types of Yukawa interactions in the THDM. 
Among the four types of Yukawa interactions, so-called Type-II and
Type-X~\cite{Ref:2hdm} deserve many interests as an effective theory of
the Higgs sector in the new physics model.
For example, the Type-II THDM is known to be the Higgs sector in the
minimal supersymmetric extension of the SM (MSSM)~\cite{Ref:HK,Ref:Djouadi2},
where one Higgs doublet couples with down-type quarks and charged
leptons and the other with up-type quarks. 
On the other hand, the Type-X THDM, in which one Higgs doublet couples
with quarks and the other with charged leptons, may be motivated by some
sort of new physics models concerning phenomena relevant to leptons and Higgs bosons,
such as tiny neutrino masses~\cite{Ref:U1X,Ref:AKS}, the positron cosmic ray
anomaly~\cite{Ref:GHK}, the Fermi-LAT gamma ray line data~\cite{Ref:BBES}, 
the muon anomalous magnetic moment~\cite{Ref:CWWY}, etc. 

% THDM and data
Verification of the THDM by using the collider data and the flavor data
has been an important task, while no positive evidence has been found so far. 
The results give constraints on the parameters in the THDM, depending
on the type of the Yukawa
interactions~\cite{Ref:mssm-HA-atlas,Ref:mssm-HA-cms,Ref:2HDM-LHC,Ref:bsg,Ref:bsg2,Ref:btaunu,Ref:tau}.
Some of them are not constrained so strongly because of the small
couplings of additional Higgs bosons with quarks, allowing a relatively light
mass of extra Higgs bosons~\cite{Ref:TypeX,Ref:AKTY,Ref:Su,Ref:Logan}.
Further studies will be continued at the upgraded LHC with
$\sqrt{s}=14$~TeV, where the discovery of more heavier particles may be
expected. 

% In this paper
On the other hand, evidences of the THDM can be probed through the 
measurement of the SM-like Higgs boson, since the coupling
constants of the SM-like Higgs boson can deviate from those in the
SM.
This is quite a realistic situation, since the Higgs couplings can be
measured very precisely, a few percent level, at the International
Linear Collider (ILC)~\cite{Ref:h-BR,Ref:ilc-TDR,Ref:ilc-Peskin}.
It may be problematic that it is not straightforward to identify the
model of new physics from such measurements, since the effects are
indirect and some models may bring the similar effects. 
To resolve this problem, one needs to combine measurements of various
observables and perform fingerprinting of the models which predict
different patterns of the deviations in the various observables.

In this paper, we focus on the determination of $\tan\beta$, the ratio
 of vacuum expectation values of the doublets in the THDMs, by using
 future precision measurements of the SM-like Higgs boson at linear
 colliders.
So far, the methods to determine $\tan\beta$ have been discussed using
 the heavy extra Higgs bosons within the context of the
 MSSM~\cite{Ref:TanB}.
However, the methods using the heavy extra bosons must follow the
 discovery of them.
Thus, these are applicable to the cases with relatively small masses,
 where already strong constraints are obtained in some types of the
 THDM~\cite{Ref:mssm-HA-atlas,Ref:mssm-HA-cms,Ref:mssm-H+}. 
On the other hand, we propose a new method to determine $\tan\beta$,
 through the branching ratios of the SM-like Higgs boson, which could be
 performed even when the discovery of extra Higgs bosons is not
 accomplished.
Our method is applicable when there exists a deviation in the gauge
 couplings of the SM-like Higgs boson.
In the general THDM, the deviation can be larger than that in the MSSM,
 since $\tan\beta$ is independent of the masses of the extra Higgs
 bosons.  
Thus, it is meaningful to investigate the $\tan\beta$ measurement in
 various situations in the masses of extra Higgs bosons. 
We evaluate the uncertainties of the $\tan\beta$ determination in our
 method in the general THDM, and compare them with those of the methods
 proposed previously.

% Organization
This paper is organized as follows.
In Section II, we give a brief review on the THDM to specify the
notation and define the parameters relevant in our study.
In Section III, three methods for the $\tan\beta$ determination are
introduced; i) the branching ratio measurement, ii) the total width
measurement of the extra Higgs bosons, and iii) the precision
measurement of the decay branching ratios of the SM-like Higgs boson. 
We apply these methods to the Type-II and Type-X THDMs. 
The simulation details are summarized in Appendix~A.
Conclusion and discussion are given in Section IV.

%%%%%%%%%%%%%%%%%%%%%%%%%%%%%%%%%%%%%%%%%%%%%%%%%%%%%%%%%%%
\section{The Two Higgs Doublet Model}

In the THDM, the SU(2) doublet scalar fields with a hypercharge $Y=1/2$
are parametrized as
%-------------------------------------------------------------------------------
\begin{align}
\Phi_i=\begin{pmatrix}i\,\omega_i^+\\\frac1{\sqrt2}(v_i+h_i-i\,z_i)
\end{pmatrix},
\end{align}
%-------------------------------------------------------------------------------
where $i=1, 2$.
The mass eigenstates are defined by introducing the mixing angles,
$\alpha$ and $\beta$, as
\begin{align}
%-------------------------------------------------------------------------------
\begin{pmatrix}h_1\\h_2\end{pmatrix}=
\begin{pmatrix}\cos\alpha&-\sin\alpha \\
\sin\alpha&\cos\alpha\end{pmatrix}
\begin{pmatrix}H\\h\end{pmatrix},\,
%-------------------------------------------------------------------------------
\begin{pmatrix}z_1\\z_2\end{pmatrix}=
\begin{pmatrix}\cos\beta&-\sin\beta \\
\sin\beta&\cos\beta\end{pmatrix}
 \begin{pmatrix}z\\A\end{pmatrix},\,
%-------------------------------------------------------------------------------
\begin{pmatrix}\omega_1^+\\\omega_2^+\end{pmatrix}=
\begin{pmatrix}\cos\beta&-\sin\beta \\
\sin\beta&\cos\beta\end{pmatrix}
\begin{pmatrix}\omega^+\\H^+\end{pmatrix},
%-------------------------------------------------------------------------------
\end{align}
where $\beta$ satisfies $\tan\beta=v_2/v_1$.
For simplicity, we assume CP conservation in the Higgs sector.
Then, there are five physical Higgs bosons, which are two CP-even
states $(h, H)$, one CP-odd state $A$, and a pair of the charged
states $H^\pm$. 
The electroweak Nambu-Goldstone bosons, $z, \omega^\pm$, are absorbed
into the weak gauge bosons. 
For the details of the Higgs potential in the THDM, see, e.g.,
Ref.~\cite{Ref:KOSY}. 

Gauge interactions of Higgs bosons in the THDM are given by normalizing
them with those in the SM as 
\begin{align}
%------------------------------------------------------------------------------- 
\frac{g^{\rm THDM}_{hVV}}{g^{\rm SM}_{hVV}} = \sin(\beta-\alpha),
 \quad 
 \frac{g^{\rm THDM}_{HVV}}{g^{\rm SM}_{hVV}} = \cos(\beta-\alpha), 
%-------------------------------------------------------------------------------
\end{align}
for $V=Z,W$. 
Thus, when $\sin(\beta-\alpha)=1$, which is so-called  ``the SM-like
limit'', $h$ has the same gauge interaction as the SM Higgs boson. 
We note that the deviation from the SM-like limit is theoretically
restricted for heavy $H$ and $A$ in the general THDM~\cite{Ref:Uni-2hdm}.
Indeed, for $m_h^2 \ll m_H^2, M^2$ with large $\tan\beta$,
$\sin(\beta-\alpha)$ can be written as $\sin^2(\beta-\alpha)\simeq
1+\lambda_1 v^2/(m_H^2\tan^2\beta)$ with $\lambda_1$ being a coefficient
of the quartic term in the potential.
A large value of $\lambda_1$ is constrained by requiring the validity
of the perturbative calculation (so-called unitarity
bound)~\cite{Ref:Uni-2hdm}.
Thus, the deviation from the SM-like limit cannot be large for heavy $H$
and $A$.

Under the $Z_2$ symmetry, there are four types of the $Z_2$ parity
assignment to the SM fermions, as listed in TABLE~\ref{Tab:type}. 
\begin{table}[t]
%-------------------------------------------------------------------------------
\begin{center}
\begin{tabular}{|c||c|c|c|c|c|c|}
\hline & $\Phi_1$ & $\Phi_2$ & $u_R^{}$ & $d_R^{}$ & $\ell_R^{}$ &
 $Q_L$, $L_L$ \\  \hline
Type-I  & $+$ & $-$ & $-$ & $-$ & $-$ & $+$ \\
Type-II & $+$ & $-$ & $-$ & $+$ & $+$ & $+$ \\
Type-X  & $+$ & $-$ & $-$ & $-$ & $+$ & $+$ \\
Type-Y  & $+$ & $-$ & $-$ & $+$ & $-$ & $+$ \\
\hline
\end{tabular}
\end{center}
\caption{Parity assignments under the softly broken $Z_2$
 symmetry~\cite{Ref:AKTY}.} 
 \label{Tab:type}
%-------------------------------------------------------------------------------
\end{table}
Then, Yukawa interactions of the SM fermions to the Higgs bosons are
given by
\begin{align}
%-------------------------------------------------------------------------------
{\mathcal L}_\text{Yukawa}^\text{THDM} =
&-{\overline Q}_LY_u\widetilde{\Phi}_uu_R^{}
-{\overline Q}_LY_d\Phi_dd_R^{}
-{\overline L}_LY_\ell\Phi_\ell \ell_R^{}+\text{H.c.},
%-------------------------------------------------------------------------------
\end{align}
where $\Phi_f$ ($f=u,d$ or $\ell$) is selected from $\Phi_1$ or $\Phi_2$
to make each vertex $Z_2$-invariant.
In terms of the mass eigenstates, the Yukawa interactions are expressed
as 
\begin{align}
%-------------------------------------------------------------------------------
{\mathcal L}_\text{Yukawa}^\text{THDM} =
&-\sum_{f=u,d,\ell} \Bigl[
+\frac{m_f}{v}\, \xi_h^f\, {\overline f}fh
+\frac{m_f}{v}\, \xi_H^f\, {\overline f}fH
-i\frac{m_f}{v}\, \xi_A^f\, {\overline f}\gamma_5fA
\Bigr] \nonumber\\
&-\Bigl\{ +\frac{\sqrt2V_{ud}}{v}\, \overline{u}
\bigl[ +m_u\, \xi_A^u\, \text{P}_L+m_d\, \xi_A^d\, \text{P}_R\bigr]d\,H^+
+\frac{\sqrt2\, m_\ell}{v}\xi_A^\ell\,\overline{\nu_L^{}}\ell_R^{}H^+
+\text{H.c.} \Bigr\},\label{Eq:Yukawa}
%-------------------------------------------------------------------------------
\end{align}
where $P_{L(R)}$ are projection operators for left-(right-)handed fermions. 
The scaling factors $\xi_\phi^f$ ($\phi=h,H,A$) are listed in
TABLE~\ref{Tab:ScalingFactor}. 
\begin{table}[t]
%-------------------------------------------------------------------------------
\begin{center}
\begin{tabular}{|c||c|c|c|c|c|c|c|c|c|}
\hline
& $\xi_h^u$ & $\xi_h^d$ & $\xi_h^\ell$
& $\xi_H^u$ & $\xi_H^d$ & $\xi_H^\ell$
& $\xi_A^u$ & $\xi_A^d$ & $\xi_A^\ell$ \\ \hline
Type-I
& $c_\alpha/s_\beta$ & $c_\alpha/s_\beta$ & $c_\alpha/s_\beta$
& $s_\alpha/s_\beta$ & $s_\alpha/s_\beta$ & $s_\alpha/s_\beta$
& $\cot\beta$ & $-\cot\beta$ & $-\cot\beta$ \\
Type-II
& $c_\alpha/s_\beta$ & $-s_\alpha/c_\beta$ & $-s_\alpha/c_\beta$
& $s_\alpha/s_\beta$ & $c_\alpha/c_\beta$ & $c_\alpha/c_\beta$
& $\cot\beta$ & $\tan\beta$ & $\tan\beta$ \\
Type-X
& $c_\alpha/s_\beta$ & $c_\alpha/s_\beta$ & $-s_\alpha/c_\beta$
& $s_\alpha/s_\beta$ & $s_\alpha/s_\beta$ & $c_\alpha/c_\beta$
& $\cot\beta$ & $-\cot\beta$ & $\tan\beta$ \\
Type-Y
& $c_\alpha/s_\beta$ & $-s_\alpha/c_\beta$ & $c_\alpha/s_\beta$
& $s_\alpha/s_\beta$ & $c_\alpha/c_\beta$ & $s_\alpha/s_\beta$
& $\cot\beta$ & $\tan\beta$ & $-\cot\beta$ \\
\hline
\end{tabular}
\end{center}
\caption{The scaling factors in each type of Yukawa interactions in
 Eq.~\eqref{Eq:Yukawa}~\cite{Ref:AKTY}.} \label{Tab:ScalingFactor}
%-------------------------------------------------------------------------------
\end{table}
Corrections to the Yukawa coupling constants of $h$ are 
$\xi^{f}_h=\sin(\beta-\alpha)+\cot\beta\cdot\cos(\beta-\alpha)$ for
$f$=$u$ in Type-II and $f$=$u,d$ in Type-X, and 
$\xi^{f}_{h}=\sin(\beta-\alpha)-\tan\beta\cdot\cos(\beta-\alpha)$ for
$f$=$d,\ell$ in Type-II and $f$=$\ell$ in Type-X. 
Thus, in the SM-like limit, the $\tan\beta$ dependence disappears in
$\xi^f_{h}$, and the Yukawa interactions of $h$ reduce to those in the
SM as well.
Otherwise, there is a $\tan\beta$ dependence in $\xi^f_{h}$. 
For the Yukawa coupling constants of $H$ and $A$, these depend significantly
on $\tan\beta$ around the SM-like limit. 

\begin{figure}[t]
%----------------------------------------------------------------------------
 \centering
 \includegraphics[height=5.2cm]{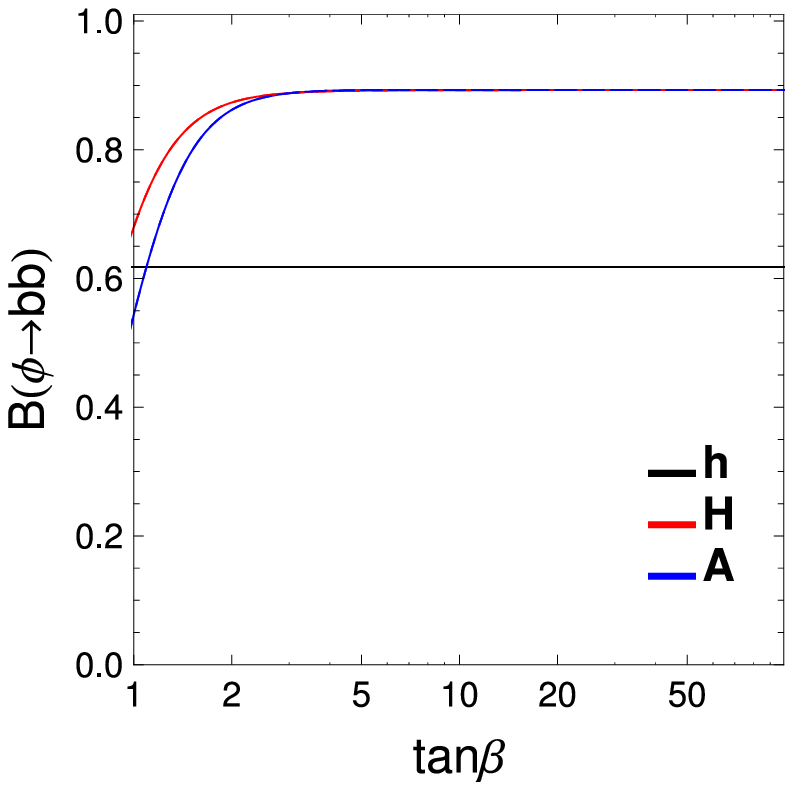} 
 \includegraphics[height=5.2cm]{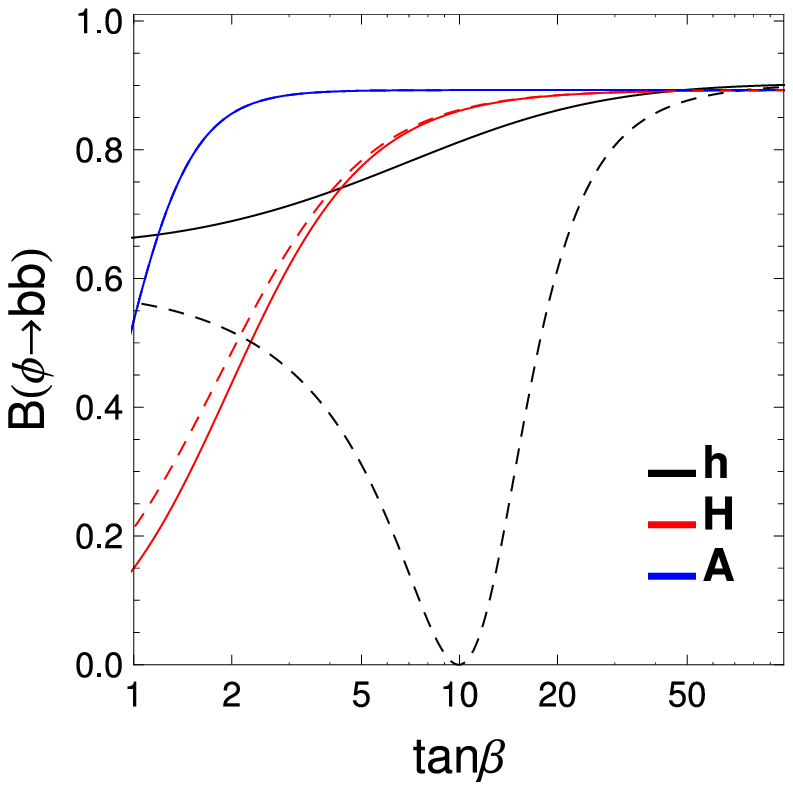} 
 \includegraphics[height=5.2cm]{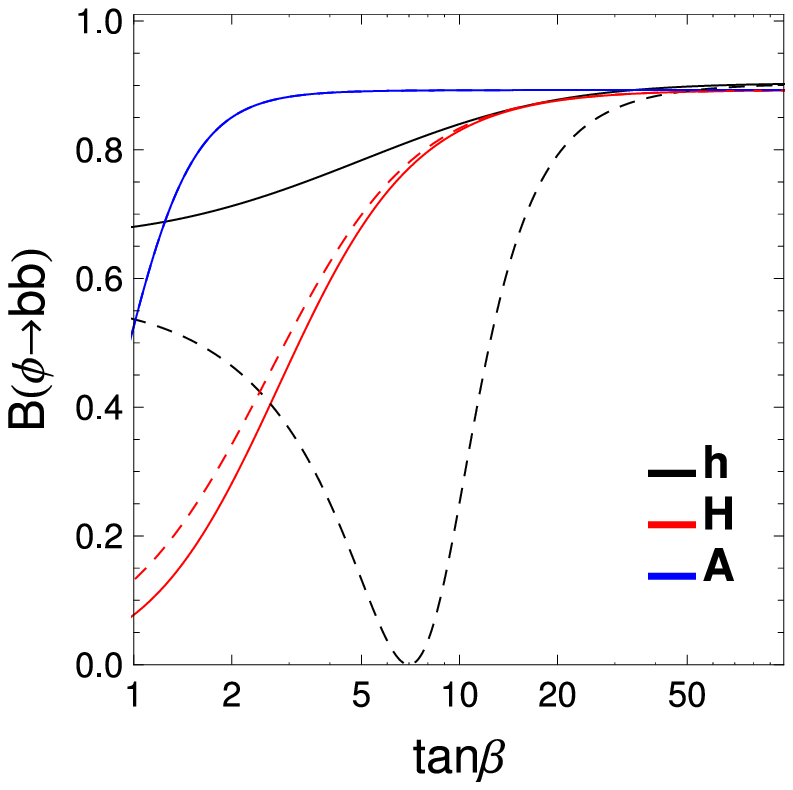} 
 \caption{The decay branching ratios are shown as a function of
 $\tan\beta$ for a fixed $\sin^2(\beta-\alpha)$ for $h\to b\bar
 b$ (black curves), $H\to b\bar b$ (red curves), and $A\to b\bar
 b$ (blue curves) decays in the Type-II THDM. 
 From left to right, $\sin^2(\beta-\alpha)$ is taken to be $1$, $0.99$,
 and $0.98$. 
 The solid (dashed) curves denote the case with $\cos(\beta-\alpha) \le 0$
 ($\cos(\beta-\alpha) \ge 0$). 
 }
 \label{FIG:Bbb}
%----------------------------------------------------------------------------
\end{figure}

The $\tan\beta$ dependence in the Yukawa coupling constants can be seen
in the branching ratios of the Higgs bosons~\cite{Ref:AKTY}.
In the Type-II and Type-X THDMs with large $\tan\beta$, a decay of $H$ and $A$
into $b\bar b$ and $\tau\tau$ is expected to be dominant, respectively. 
In FIG.~\ref{FIG:Bbb}, we evaluate the $\tan\beta$ dependence in the
branching ratios of $H,A$ and also $h$ into $b\bar b$ in the Type-II
THDM.
The three panels correspond to the cases with $\sin^2(\beta-\alpha)=1$
(left), 0.99 (middle) and 0.98 (right), 
and the case with $\cos(\beta-\alpha) \le 0$ ($\cos(\beta-\alpha) \ge
0$) is plotted in the solid (dashed) curves. 
For each panel, ${\mathcal B}^{\phi}_{bb}\equiv{\mathcal B}(\phi\to
b\bar b)$ for $\phi=h$, $H$ and $A$ are plotted in black, red and blue
curves respectively.
Here, the masses of $H$ and $A$ are taken commonly to be $200$ 
GeV.\footnote{
The mass of the charged Higgs boson is also assumed to be $200$ GeV, 
in order to avoid a severe constraint from the $\rho$ parameter data~\cite{Ref:rho-2hdm,Ref:rho2-2hdm,Ref:rho3-2hdm,Ref:KOTT}
} 
The branching ratios ${\mathcal B}_{bb}$ for $H$ and $A$ grow with
$\tan\beta$, and reach a saturation point above which the values are
fixed to ${\mathcal B}_{bb}\simeq 0.9$ (and the rest is ${\mathcal 
B}_{\tau\tau}\simeq 0.1$). 
A slightly large $\sin(\beta-\alpha)$ dependence in ${\mathcal B}^H_{bb}$
comes from the $H\to W^+W^-$ and $ZZ$ decay modes which rapidly 
increases with $\cos^2(\beta-\alpha)$.
On the other hand, for ${\mathcal B}^h_{bb}$, there is no $\tan\beta$
dependence in the SM-like limit. 
However, once $\sin(\beta-\alpha)$ deviates from unity, ${\mathcal
B}^h_{bb}$ shows a significant $\tan\beta$ dependence, with a
large difference by the sign of $\cos(\beta-\alpha)$. 
Thus, the deviation from the SM-like limit, $\sin(\beta-\alpha)-1$,
triggers the $\tan\beta$ dependence in ${\mathcal B}^h_{bb}$. 
We note that $\sin^2(\beta-\alpha)$ should be measured very accurately by a few
percent~\cite{Ref:Zh-lep}, by using the cross section measurement of the
$e^+e^-\to Zh$ process at the ILC. 
On the other hand, the determination of the sign of $\cos(\beta-\alpha)$
is not straightforward.
In the following discussion, we present the analysis for fixed 
$\sin(\beta-\alpha)$ values in the cases of a positive and negative
sign of $\cos(\beta-\alpha)$.
We note that $\cos(\beta-\alpha)<0$ is derived in the MSSM.

\begin{figure}[tb]
%----------------------------------------------------------------------------
 \centering
 \includegraphics[height=5.2cm]{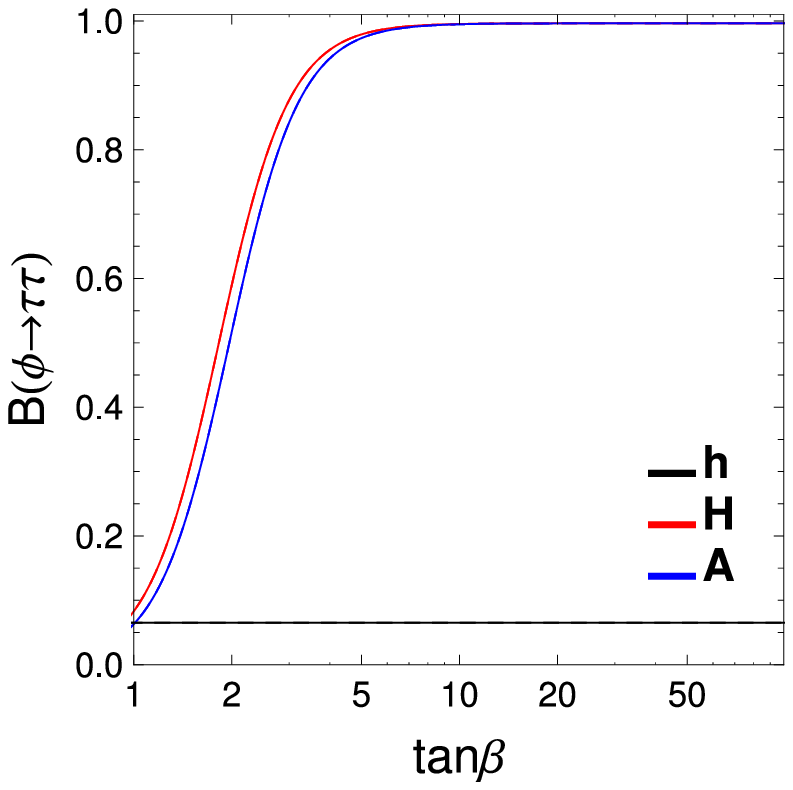} 
 \includegraphics[height=5.2cm]{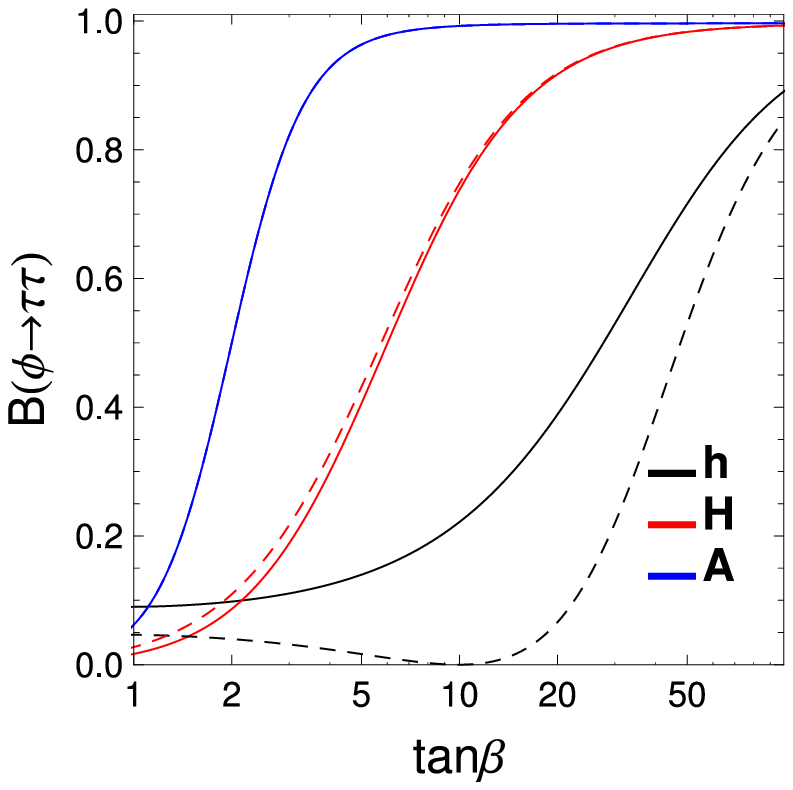} 
 \includegraphics[height=5.2cm]{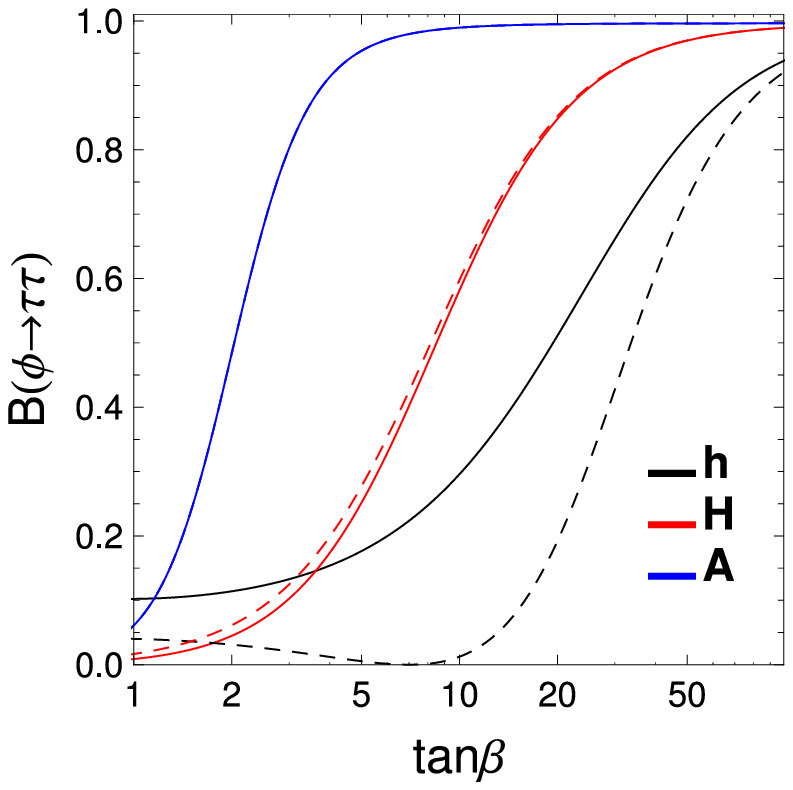} 
 \caption{The decay branching ratios are shown as a function of
 $\tan\beta$ with a fixed value of $\sin^2(\beta-\alpha)$ for $h\to
 \tau\tau$ (black curves), for $H\to \tau\tau$ (red curves), and
 for $A\to \tau\tau$ (blue curves) in the Type-X THDM.
 From left to right, $\sin^2(\beta-\alpha)$ is taken to be $1$, $0.99$,
 and $0.98$. 
 The solid (dashed) curves denote the case with $\cos(\beta-\alpha) \le
 0$ ($\cos(\beta-\alpha) \ge 0$). 
 }
 \label{FIG:Btautau}
%----------------------------------------------------------------------------
\end{figure}

In FIG.~\ref{FIG:Btautau}, $\tan\beta$ dependences in the
branching ratios for $h\to\tau\tau$ (black curves), 
$H\to\tau\tau$ (red curves) and $A\to\tau\tau$ (blue curves)
decays in the Type-X THDM are plotted, where the results of 
$\sin^2(\beta-\alpha)=1$ (left panel), 0.99 (middle) and 0.98 (right)
with $\cos(\beta-\alpha) \le 0$ (solid curves) and $\cos(\beta-\alpha)
\ge 0$ (dashed) are considered. 
The masses of $H$ and $A$ are fixed to be $200$ GeV. 
The qualitative features are almost similar with those in the Type-II. 
The branching ratios reach a saturation point close to unity at a
rather large $\tan\beta$ value. 

%%%%%%%%%%%%%%%%%%%%%%%%%%%%%%%%%%%%%%%%%%%%%%%%%%%%%%%%%%%
\section{The $\tan\beta$ determination}

In this section, we investigate the methods for the determination of
$\tan\beta$ in the THDM at linear colliders. 
In Ref.~\cite{Ref:TanB}, methods by using the production and decays
of $H$ and $A$ at linear colliders are studied in a context of the MSSM.
In addition, we propose to utilize the precise measurement of the decay
branching ratios of $h$, and compare its sensitivity with those of
the previous methods in Ref.~\cite{Ref:TanB}.
Then, we calculate the accuracy of the determination of $\tan\beta$ in
the Type-II and Type-X THDMs by three methods which are described as
follows: 
\\

(i) The first method is based on the measurement of the branching ratios of
$H$ and $A$ in the $e^+e^-\to HA$ process~\cite{Ref:TanB}.
Since the masses of the neutral Higgs bosons can be measured by the
invariant mass distributions in an appropriate decay mode, the branching
ratios can be predicted as a function of $\tan\beta$.
Thus, $\tan\beta$ can be determined by measuring the decay branching
ratios of $H$ and $A$.
Because the $\tan\beta$ dependence in the branching ratios is large in
the relatively small $\tan\beta$ regions, as we see in
Figs.~\ref{FIG:Bbb} and \ref{FIG:Btautau}, the method is useful for those
regions. 

(ii) The second method is based on the measurement of the total decay
widths of $H$ and $A$~\cite{Ref:TanB}. 
For large $\tan\beta$, the total decay widths of $H$ and $A$ are governed by the
$b\bar b~$ and $\tau\tau$ decay modes in the Type-II and Type-X THDMs, 
respectively, whose partial decay widths are proportional to $(\tan\beta)^2$.
If the total decay widths are wider than the detector resolution for the
invariant mass measurement, we can directly measure the absolute value 
of the total decay widths.
Thus, $\tan\beta$ can be extracted from the total decay widths 
in the large $\tan\beta$ regions.

(iii) In addition to these two methods, we propose to utilize the
precision measurement of the decay branching ratios of $h$.
For $m_h=125$~GeV, the main decay modes of $h$ are expected to be
measured precisely by a few percent level at the ILC~\cite{Ref:h-BR}.
In the THDM, the decay branching ratios of $h$ depend on $\tan\beta$, as
long as $\sin(\beta-\alpha)<1$ which can be determined independently. 
This means that the precision measurement of the decay of $h$ can probe
$\tan\beta$. 

In the following subsections, we show the detailed analysis of these
methods in the Type-II and Type-X THDMs.  
We also comment on the cases for the other types in the THDM.

\subsection{Sensitivity to $\tan\beta$ in the Type-II THDM}

In this subsection, we present our numerical analysis for the
sensitivities of the $\tan\beta$ measurements in the Type-II THDM.
The same analysis for the Type-X THDM is given in the next subsection.
 
Following Ref.~\cite{Ref:TanB}, the $1\sigma$ sensitivity to $\tan\beta$
from the measurement of the branching ratios of $H$ and $A$ is defined by
$N(\tan\beta\pm\Delta\tan\beta)=N_{\rm obs}\pm\sqrt{N_{\rm obs}}$ with
$N(\tan\beta)=\sigma^{}_{HA}\cdot{\mathcal
B}^H_{bb}(\tan\beta)\cdot{\mathcal B}^A_{bb}(\tan\beta)\cdot{\mathcal
L}_{\rm int}\cdot\epsilon_{4b}$, where $\sigma^{}_{HA}$ is the $HA$
production cross section, ${\mathcal L}_{\rm int}$ is the integrated
luminosity, and $N_{\rm obs}$ is the number of the $4b$ signal events
after the selection cuts.
The production cross section and the number of the signal events are
evaluated for $m_H=m_A=200$~GeV with $\sqrt{s}=500$~GeV and ${\mathcal
L}_{\rm int}=250$~fb$^{-1}$. 
The acceptance ratio $\epsilon_{4b}$ of the $4b$ final
states in the signal process of $HA$ production is set to be 50\% (see
Appendix~A for details). 

For the width measurement of $H$ and $A$, the detector resolution for
the Breit-Wigner width of the invariant mass distribution of $b\bar b$
($M_{bb}$) is taken to be $\Gamma_\text{res} = 11$~GeV with the 10\%
systematic error.
In order to reduce the combinatorial uncertainty due to the $4b$ final
state, the signal events are chosen around the central peak regions in the
invariant mass distribution.
This selection efficiency is estimated to be 40\% for $M_{bb} \pm 10$
GeV. 
The width to be observed is $\Gamma^R_{H/A} =
\frac12[\sqrt{(\Gamma_\text{tot}^H)^2+(\Gamma_\text{res})^2} +
\sqrt{(\Gamma_\text{tot}^A)^2+(\Gamma_\text{res})^2}]$~\cite{Ref:TanB},
and the $1\sigma$ uncertainty is given by
$\Delta\Gamma^R_{H/A}=[(\Gamma^R_{H/A}/\sqrt{2N_{\rm obs}})^2 +
(\Delta\Gamma_\text{res}^\text{sys})^2]^{1/2}$, where $N_{\rm obs}$ 
is the number of the events after the selection cuts and the invariant
mass cut. 
We then obtain the $1\sigma$ sensitivity for the $\tan\beta$ determination
by $\Gamma_{H/A}(\tan\beta \pm \Delta\tan\beta)
=\Gamma^R_{H/A}\pm\Delta\Gamma^R_{H/A}$, where
$\Gamma_{H/A}(\tan\beta)=\frac12[\Gamma_{H}(\tan\beta)+\Gamma_{A}(\tan\beta)]$.

For the $\tan\beta$ determination by using the decay branching ratio of
$h$, we evaluate the sensitivity to $\tan\beta$ from the uncertainties
of the ${\mathcal B}^h_{bb}$ measurement. 
The $\tan\beta$ sensitivity is obtained by solving ${\mathcal
B}^h_{bb}(\tan\beta \pm \Delta\tan\beta) = {\mathcal B}^h_{bb} \pm
\Delta{\mathcal B}^h_{bb}$, where the accuracy is evaluated from the
simulation result for that in the SM case by rescaling the statistical
factor by taking into account the changes of the production cross section as
well as the branching ratio. 
The reference points of $\Delta{\mathcal B}^h_{bb}/{\mathcal B}^h_{bb}$
in the SM case are $1.3\%$ ($1\sigma$) and $2.7\%$ ($2\sigma$), 
which are estimated in the recent simulation study for
$\sqrt{s}=250$~GeV and ${\mathcal L}_{\rm int}=250$~fb$^{-1}$~\cite{Ref:h-BR}.

Notice that the cross section times the branching ratio is expected to
be measured more precisely by $\Delta(\sigma_{Zh}{\mathcal
B}^h_{bb})/(\sigma_{Zh}{\mathcal B}^h_{bb})=1\%$ at the $2\sigma$
confidence level (CL).~\cite{Ref:h-BR}. 
However, the uncertainty of the cross section amounts to 
$\Delta\sigma_{Zh}/\sigma_{Zh} = 2.5\%$ at the $2\sigma$
CL\ by assuming the analysis of the leptonic decays
of the recoiled $Z$ boson~\cite{Ref:Zh-lep}. 
Therefore, the accuracy of the branching ratio measurement is limited by
the uncertainty of the $\sigma_{Zh}$ determination. 
If the cross section measurement is improved up to $0.8\%$ at the $2\sigma$
CL\ by using the analysis of the hadronic decays of the recoiled $Z$
boson~\cite{Ref:Zh-had,Ref:ilc-TDR}, it would give much better sensitivities to
$\tan\beta$ from the $h$ decay. 

\begin{figure}[tb]
%----------------------------------------------------------------------------
 \centering
 \includegraphics[height=5.2cm]{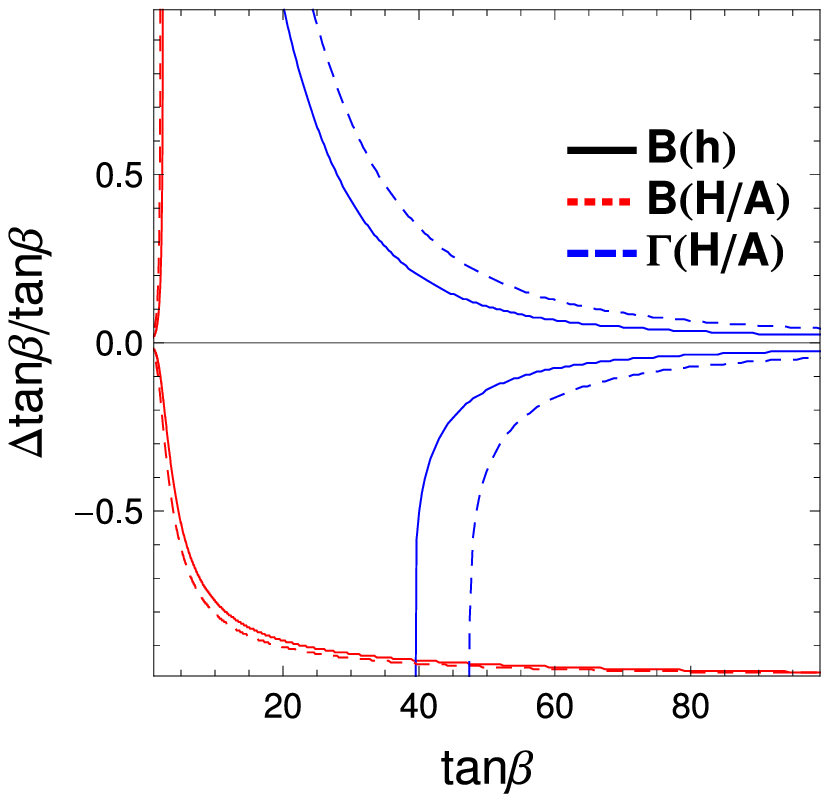} 
 \includegraphics[height=5.2cm]{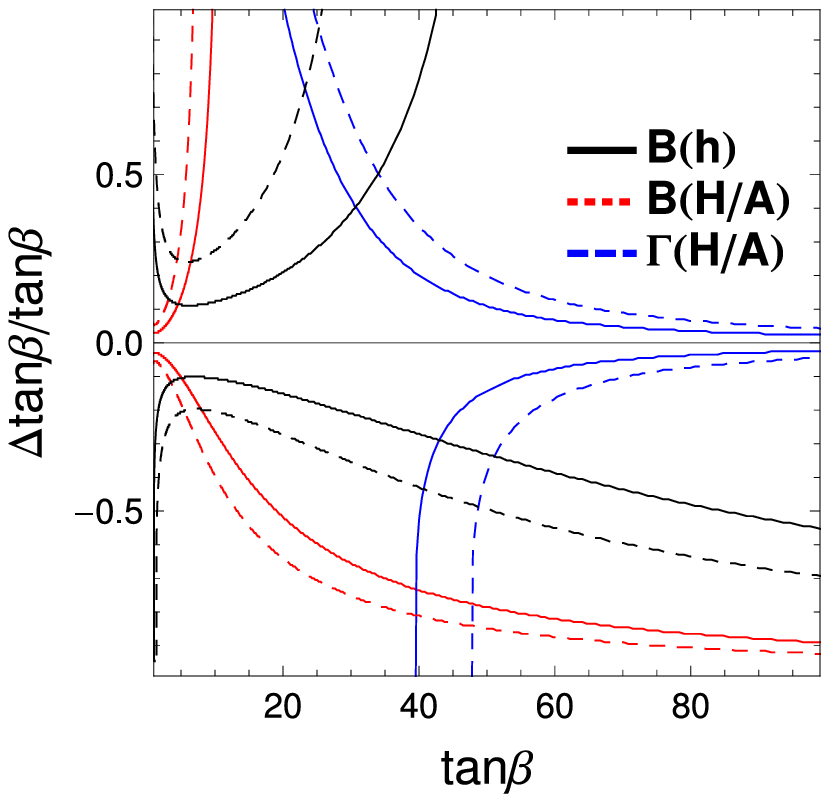} 
 \includegraphics[height=5.2cm]{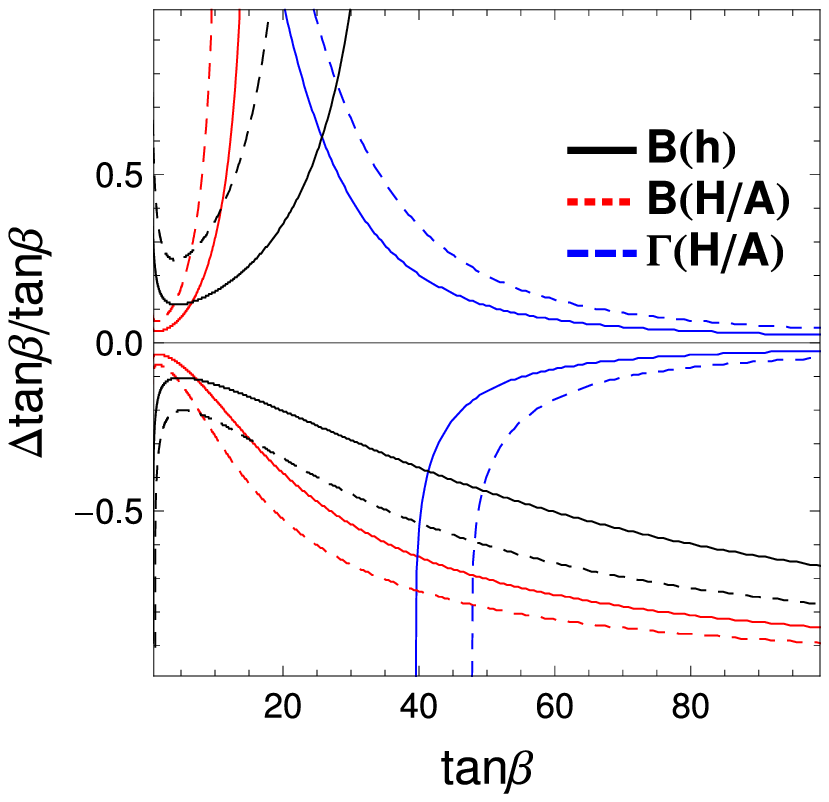} 
 \caption{Sensitivities to the $\tan\beta$ measurement by the three
 methods in the Type-II THDM.
From left to right, $\sin^2(\beta-\alpha)$ is taken to be $1$, $0.99$,
 and $0.98$, with $\cos(\beta-\alpha) \le 0$.
Estimated $\Delta\tan\beta/\tan\beta$ by using the branching ratio of
 $H/A\to b\bar b$ (red curves), the total width of $H/A$ (blue curves),
 and the branching ratio of $h\to b\bar b$ (black curves) are plotted as
 a function of $\tan\beta$.
The solid curves stand for $1\sigma$ sensitivities, and the dashed
 curves for $2\sigma$. 
For $HA$ production, $m_H^{}=m_A^{}=200$ GeV with $\sqrt{s}=500$ GeV and
${\mathcal L}_{\rm int}=250~\text{fb}^{-1}$ are assumed.
For the $h\to b\bar b$ measurement, $\Delta{\mathcal
B}/{\mathcal B} = 1.3\%$ ($1\sigma$) and $2.7\%$ ($2\sigma$)
are used.
}~\label{FIG:2HDM-II}
%----------------------------------------------------------------------------
%\end{figure}
%
%\begin{figure}[tb]
%----------------------------------------------------------------------------
\vspace{2em}
 \centering
 \includegraphics[height=5.2cm]{TypeII_dTanB_100.eps} 
 \includegraphics[height=5.2cm]{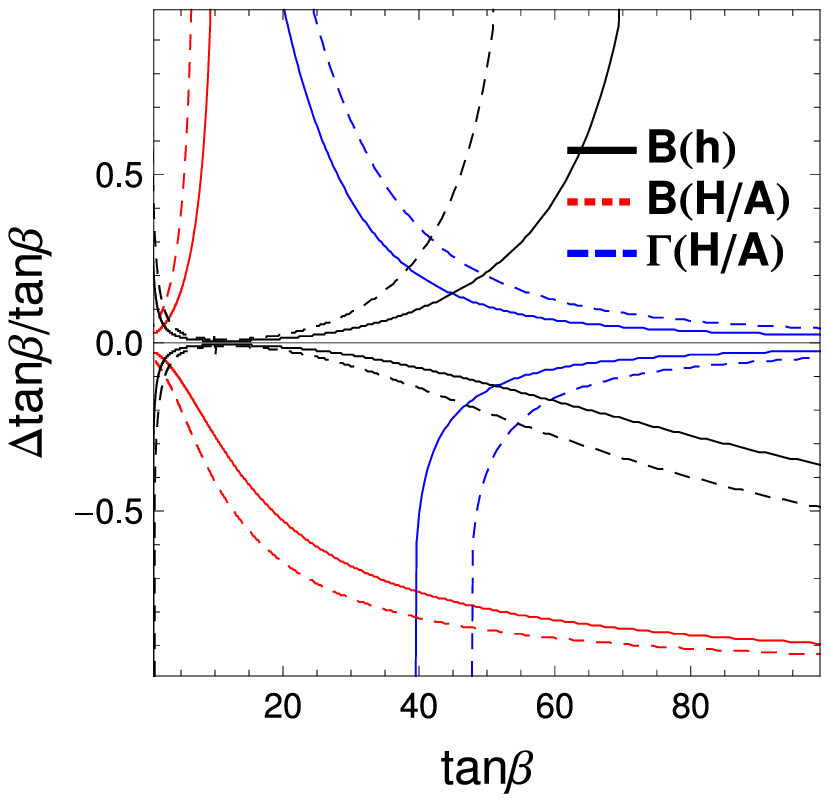} 
 \includegraphics[height=5.2cm]{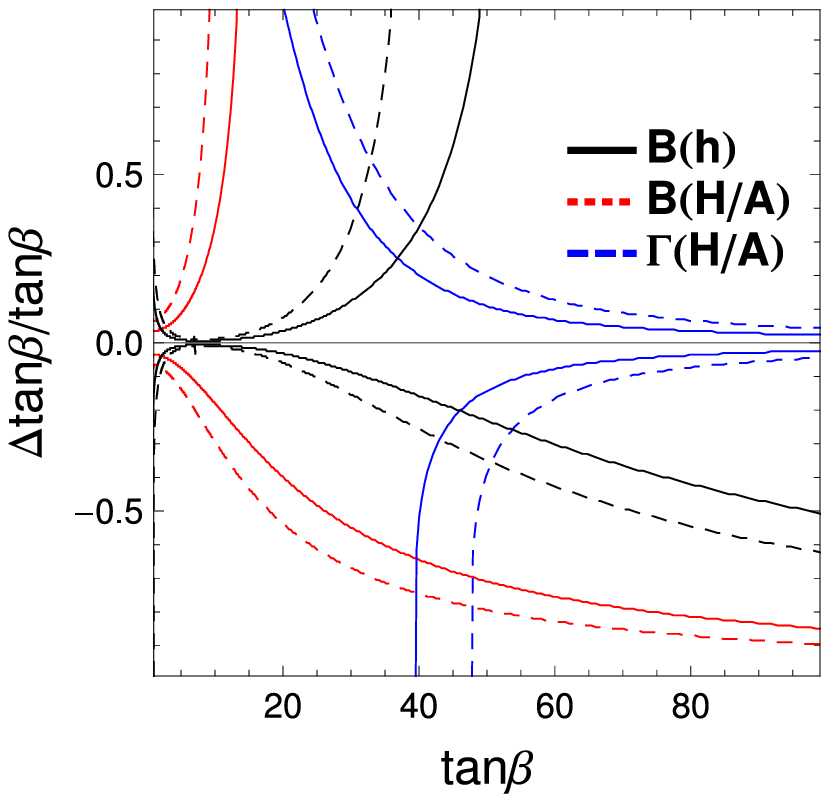} 
 \caption{The same as FIG.~\ref{FIG:2HDM-II}, but for $\cos(\beta-\alpha)
 \ge 0$.
 }
 \label{FIG:2HDM-II-posi}
%----------------------------------------------------------------------------
\end{figure}

In FIG.~\ref{FIG:2HDM-II}, our numerical results for the three methods are shown.
The results for $1\sigma$ (solid) and $2\sigma$ (dashed) sensitivities
for the branching ratios, the total width of $H$ and $A$, and the branching
ratio of $h$ are plotted in the red, blue and black curves, respectively. 
The parameter $\sin^2(\beta-\alpha)$ is set to be $1$ (left), 
$0.99$ (middle), and $0.98$ (right) with $\cos(\beta-\alpha) \le 0$. 
The case with $\cos(\beta-\alpha)\ge 0$ is also shown in FIG.~\ref{FIG:2HDM-II-posi}.

In the Type-II THDM, the three methods complementary cover the wide
range of $\tan\beta$ values. 
In the SM-like limit (left panels of FIGs.~\ref{FIG:2HDM-II} and
\ref{FIG:2HDM-II-posi}), the method using the $h$ decay has no
sensitivity to $\tan\beta$.
But, for $\sin^2(\beta-\alpha)=0.99$, 0.98, there are certain
$\tan\beta$ regions, from about 5 to 30$\sim$40, where it gives the best
sensitivity among the three methods.
The sensitivity for $\cos(\beta-\alpha)>0$ is better than that for
$\cos(\beta-\alpha)<0$, because the former case has a large gradient
$|d{\mathcal B}/d(\tan\beta)|$, as shown in FIG.~\ref{FIG:Bbb}.
On the other hand, in the $\cos(\beta-\alpha)>0$ case, there is a
two-fold ambiguity in determining $\tan\beta$ from the ${\mathcal
B}_{bb}^h$. 
We expect that this ambiguity is resolved by using the other methods, 
or by measuring the branching ratios in the other $h$ decay modes, such
like into $gg$ and $c\bar c$.

The sensitivity for the $h$ decay becomes worse for large $\tan\beta$,
where ${\mathcal B}^h_{bb}$ is saturated at about 90\% as shown in
FIG.~\ref{FIG:Bbb}. 
We note that, however, such a large deviation in the decay branching
ratios of $h$ should be constrained from the LHC data, where,
e.g., the prediction of ${\mathcal B}(h\to ZZ^*)$ can be different from
that in the SM. 

\subsection{Sensitivity to $\tan\beta$ in the Type-X THDM}

In the Type-X THDM, the sensitivities to $\tan\beta$ are evaluated
similarly to the case in the Type-II THDM, but the decay mode of
$\tau\tau$ is used instead of that of $b\bar b$.
For $HA$ production, the acceptance for the $4\tau$ final state is
estimated to be 50\% (for details, see Appendix A). 
The detector resolution for the Breit-Wigner width in the invariant mass
distribution of $\tau\tau$ ($M_{\tau\tau}$) is obtained with the
use of the collinear approximation~\cite{Ref:KTY}, which is estimated to
be $7$~GeV. 
The selection efficiency due to the mass window cut $M_{\tau\tau} \pm
10$~GeV is 30\%. 
For the $h\to\tau\tau$ decay, expected accuracy of the
measurement of the branching ratio at the ILC is $\Delta {\mathcal
B}^h_{\tau\tau}/{\mathcal B}^h_{\tau\tau} = 5\%$ (2\%) in the $2\sigma$
($1\sigma$) CL\ in the SM for $\sqrt{s}=250$~GeV and
${\mathcal L}_{\rm int}=250$~fb$^{-1}$~\cite{Ref:h-BR}. 
We rescale them to the case in the Type-X THDM with certain values of
$\sin^2(\beta-\alpha)$ and $\tan\beta$ taking into account the
changes of the number of the signal events.

\begin{figure}[tb]
%----------------------------------------------------------------------------
 \centering
 \includegraphics[height=5.2cm]{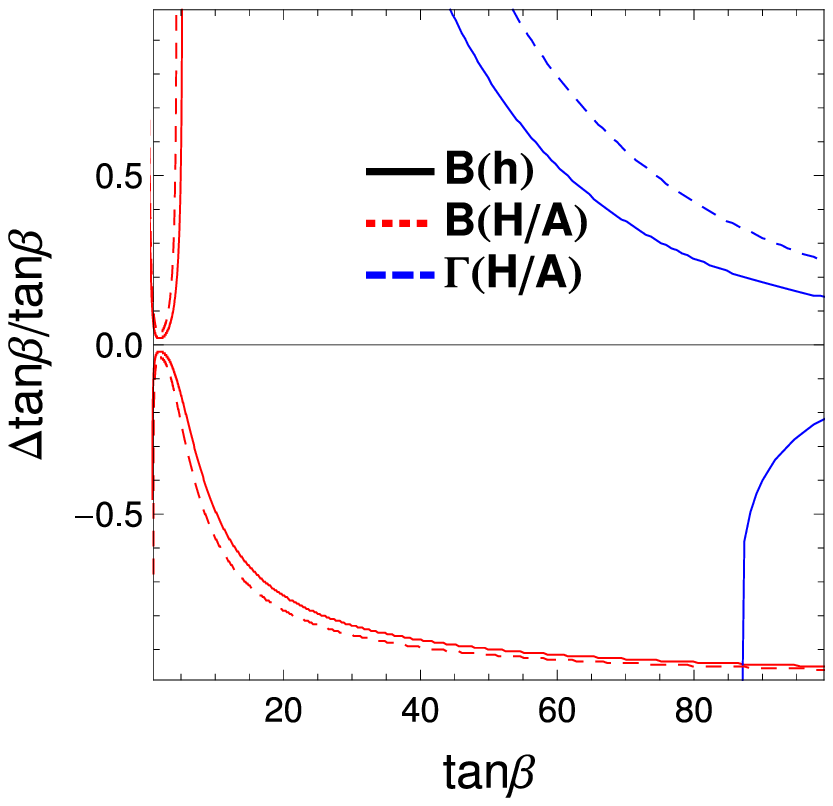} 
 \includegraphics[height=5.2cm]{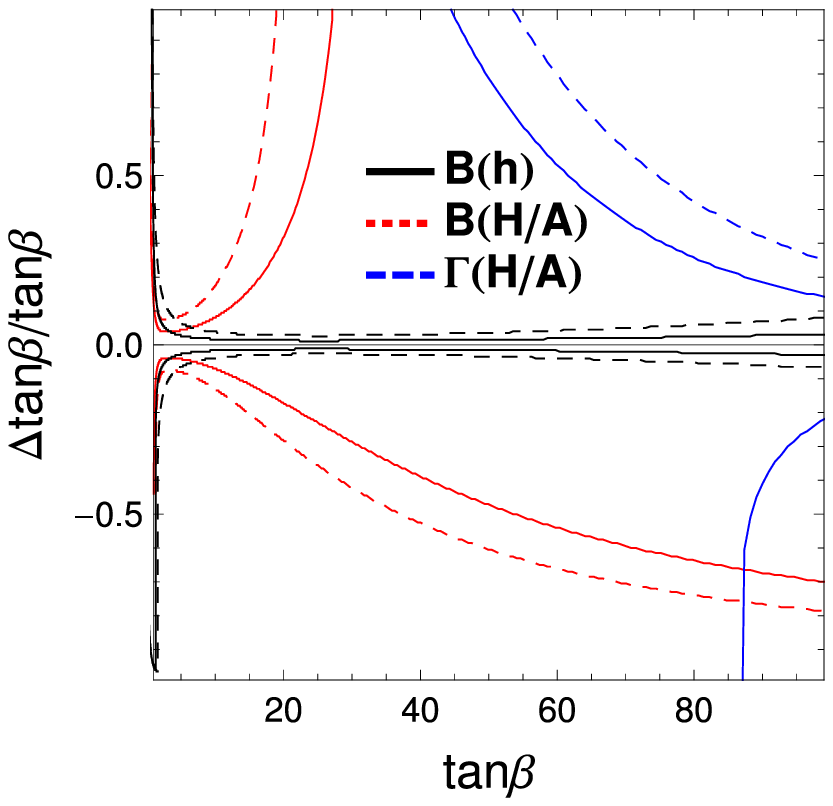} 
 \includegraphics[height=5.2cm]{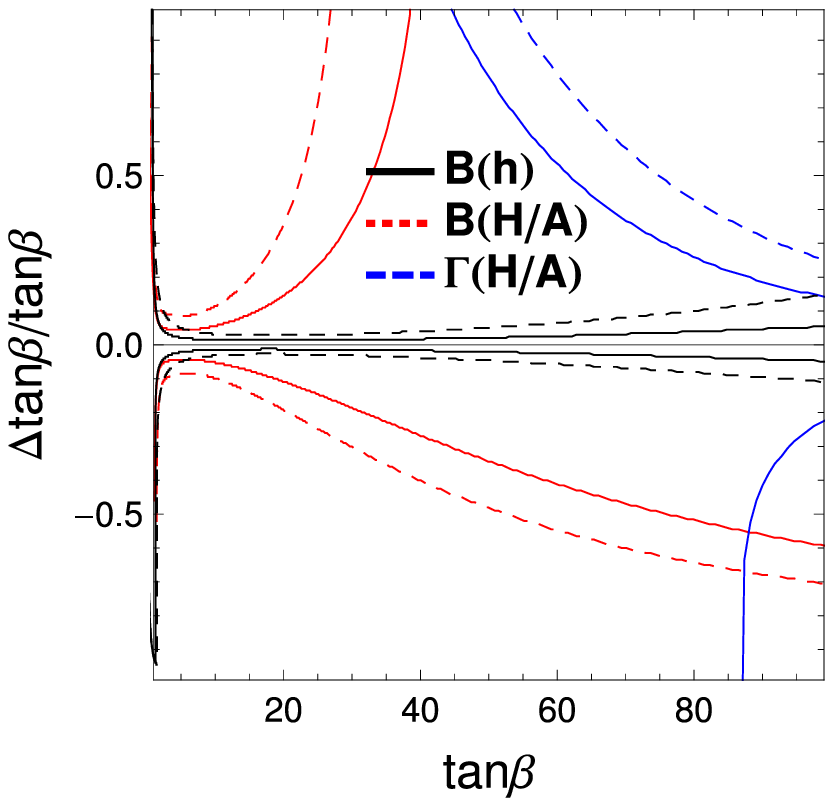} 
 \caption{The same as FIG.~\ref{FIG:2HDM-II}, but $\tau\tau$ decay modes 
 are used for the analysis in the Type-X THDM.
 From left to right, $\sin^2(\beta-\alpha)$ is taken to be $1$, $0.99$,
 and $0.98$, with $\cos(\beta-\alpha) \le 0$.
 For ${\mathcal B}^h_{\tau\tau}$, $\Delta{\mathcal B}/{\mathcal B} = 2\%
 $ ($1\sigma$) and $5\%$ ($2\sigma$) are assumed.  
 }
 \label{FIG:2HDM-X}
%----------------------------------------------------------------------------
%\end{figure}
%
%\begin{figure}[tb]
%----------------------------------------------------------------------------
\vspace{2em}
 \centering
 \includegraphics[height=5.2cm]{TypeX_dTanB_100.eps} 
 \includegraphics[height=5.2cm]{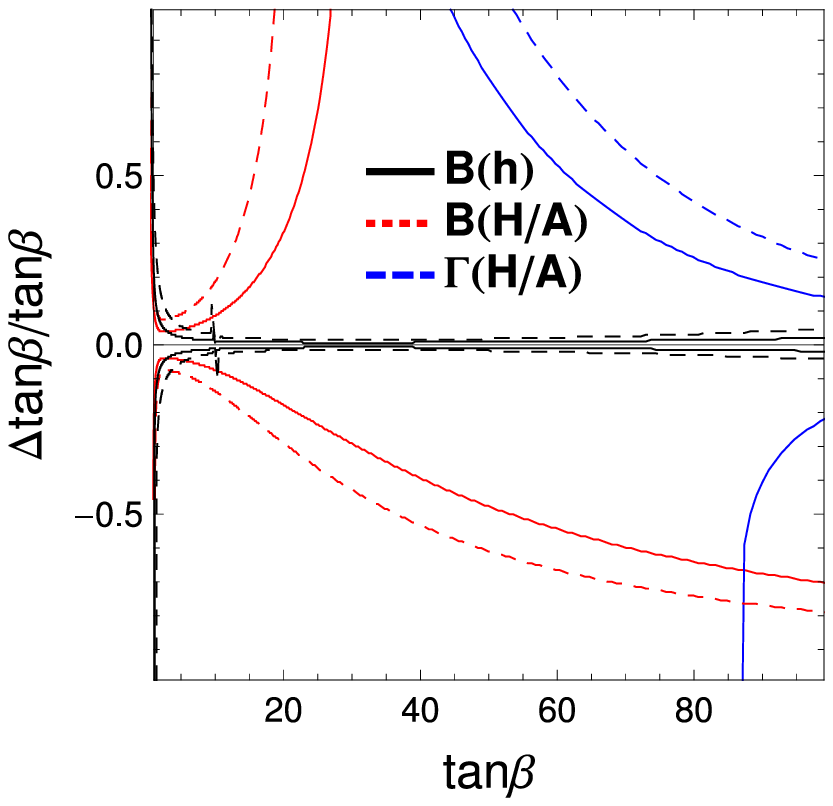} 
 \includegraphics[height=5.2cm]{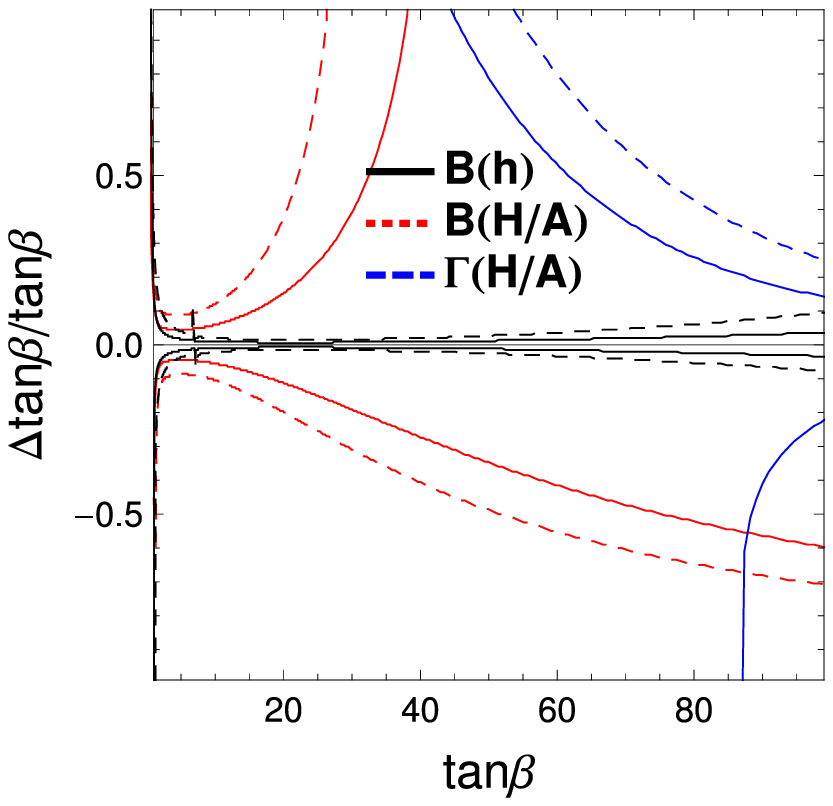} 
 \caption{The same as FIG.~\ref{FIG:2HDM-X}, but for $\cos(\beta-\alpha)
 \ge 0$.
 }
 \label{FIG:2HDM-X-posi}
%----------------------------------------------------------------------------
\end{figure}

In FIGs.~\ref{FIG:2HDM-X} and \ref{FIG:2HDM-X-posi}, the numerical
results in the Type-X THDM are presented in the same manner as in 
FIGs.~\ref{FIG:2HDM-II} and \ref{FIG:2HDM-II-posi} in the Type-II case,
respectively.
With or without the assumption of $\sin(\beta-\alpha)=1$, the total
width measurement of $H$ and $A$ is a useful probe for the large
$\tan\beta$ regions. 
For the smaller $\tan\beta$ regions, the branching ratio measurement of
$H$ and $A$ can probe $\tan\beta$ well. 
As noted in the Type-II case, the $h$ decay does not have a
sensitivity to $\tan\beta$ in the SM-like limit (left panels). 
However, for $\sin(\beta-\alpha)=0.99$ and 0.98, ${\mathcal
B}^h_{\tau\tau}$ measurement can give the best sensitivity for a
wide range of $\tan\beta$ values. %, $10\lesssim\tan\beta\lesssim100$. 
This is because the branching ratio of $h\to\tau\tau$ is about 10 times
smaller than that of $h \to b\bar b$, and thus the saturation of the
branching ratio for large $\tan\beta$ is relatively delayed as compared 
to the case in the Type-II THDM. 
This is also true for the branching ratios of $H/A\to\tau\tau$. 
In the $\cos(\beta-\alpha)>0$ case, the method using the $h$ decay 
has two-fold ambiguity to determine $\tan\beta$ from the ${\mathcal
B}_{\tau\tau}^h$ measurement.
This ambiguity is expected to be solved by using the other methods. 

\subsection{Sensitivity to $\tan\beta$ in the other types of the THDM}

Here, we comments on the $\tan\beta$ measurements for the other types in
the THDM.
In the Type-I THDM, the Yukawa coupling constants are universally
changed from those in the SM.
In the SM-like limit, $\sin(\beta-\alpha)=1$, Yukawa interactions for 
$H$ and $A$ become weak for $\tan\beta > 1$. 
As for the $\tan\beta$ measurement at the ILC, the method by using the
total width of $H$ and $A$ is useless, because the absolute value of the decay
width is too small as compared to the detector resolution.
Without the SM-like limit, the branching ratio measurement of $H$ and $A$
using the fermionic decay modes may be difficult, because the bosonic decay
modes $H\to WW$ and $A\to Zh$ become important.
Furthermore, the decays of $h$ are almost unchanged from the SM because
of no $\tan\beta$ enhancement. 
Thus, the $\tan\beta$ determination in the Type-I THDM seems to be
difficult even at the ILC. 

In the Type-Y THDM, the sensitivities to $\tan\beta$ at the ILC would be
similar to those in the Type-II THDM, because the Yukawa interactions of
the neutral scalar bosons with the bottom quarks are enhanced by
$\tan\beta$ as the same way as those in the Type-II THDM. 

%%%%%%%%%%%%%%%%%%%%%%%%%%%%%%%%%%%%%%%%%%%%%%%%%%%%%%%%%%%
\section{Conclusion and discussion}

We have studied the physics potential for the $\tan\beta$ determination
at the ILC in the THDMs.
In addition to the masses of the extra Higgs bosons, $\tan\beta$ and
$\sin(\beta-\alpha)$ are important parameters in the THDM, which
describe the electroweak symmetry breaking sector in the model.
At the ILC, the parameter $\sin^2(\beta-\alpha)$ is determined very precisely. 
The branching ratios of the $h$ decay can also be measured precisely and
independently of $\sin^2(\beta-\alpha)$. 
Since the Yukawa coupling constants of $h$ are modified if
$\sin(\beta-\alpha) \ne1$, the combination of these measurements can
constrain $\tan\beta$.
If $H$ and $A$ are light, measurements of the decay branching ratio and
the total decay width can also probe $\tan\beta$.

In this paper, we have studied the sensitivities of the $\tan\beta$
measurements using these observables in the Type-II and Type-X THDMs. 
In the Type-II THDM, the down-type quark and the lepton Yukawa
interactions of $H$ and $A$ largely depend on $\tan\beta$. 
Since the masses of $H$ and $A$ have been strongly constrained by the
LHC data and the flavor data, direct searches of them and the precision
measurements of their properties should require relatively high
collision energy at the ILC.
On the other hand, the $h$ decay can explore $\tan\beta$
if $\sin(\beta-\alpha)\ne1$ through the precision measurement of its
branching ratios.

In the Type-X THDM, only the leptonic Yukawa interactions of $H$ and $A$
are enhanced for the large $\tan\beta$ regions. 
Since they have less interaction with quarks, a severe bound from the
LHC data and the flavor data can be evaded. 
Therefore, $H$ and $A$ can be light enough to be produced at the ILC.
If they are light, $\tan\beta$ can be determined by the direct
measurement of their properties at the ILC. 
We have compared the sensitivities to $\tan\beta$ using these
measurements and the $h$ decay.
We find that the precision study of the branching ratios in the
$h$ decay is very useful to determine $\tan\beta$ in the Type-X THDM
for the wide range of parameter space. 

In conclusion, we have studied the methods of the $\tan\beta$
measurement in general THDMs at linear colliders.
In addition to the methods previously proposed by using the extra Higgs
bosons, we have discussed the method which uses the precision
measurement of the $h$ decay. 
We have found that $\tan\beta$ can be determined very well in a wide range 
of $\tan\beta$ values at the ILC by combining these methods.

%%%%%%%%%%%%%%%%%%%%%%%%%%%%%%%%%%%%%%%%%%%%%%%%%%%%%%%%%%%
\acknowledgments
 We would like to thank Kei Yagyu for collaboration in an early stage of
 this project.
This work was in part supported by Grant-in-Aid for Scientific research
from the Ministry of Education, Science, Sports, and Culture (MEXT),
Japan, Nos.\ 22244031, 23104006, 23104011, and 24340046, and the Sasakawa
Scientific Research Grant from The Japan Science Society. 

\newpage
%%%%%%%%%%%%%%%%%%%%%%%%%%%%%%%%%%%%%%%%%%%%%%%%%%%%%%%%%%%
\appendix
\section{Simulation detail in $HA$ production}

In this appendix, we present our estimation of the efficiencies
for the $4b$ and $4\tau$ events and the resolution of the
width in the $bb$ and $\tau\tau$ invariant mass distributions in the
$e^+e^-\to HA$ process. 
The simulation is performed by using the hadron-level events with {\tt
Pythia}~\cite{Ref:PYTHIA} and {\tt FastJet}~\cite{Ref:FastJet}
for jet clustering. 

For each event, we collect the final state particles with $|\eta|<2$
where the pseudorapidity is
$\eta=\frac{1}{2}\ln\frac{1+\cos\theta}{1-\cos\theta}$, and $\theta$ is
the polar angle of particle's momentum in the laboratory frame. 
For charged particles, a cut on the transverse momentum, $p_T^{}>300$~MeV,
is also applied.
The momenta are smeared by Gaussian distributions with 
$\sigma_E^{}/E=15\%/\sqrt{(E~[{\rm GeV}])}+1\%$ for photons, 
$\sigma_E^{}/E=40\%/\sqrt{(E~[{\rm GeV}])}+2\%$ for neutral hadrons, and 
$\sigma_p/p_T=10^{-4}\times(p_T^{}~[{\rm GeV}])+0.1\%$ for charged
particles, where $\sigma$'s are a dispersion of each
distribution.
Then, the particles are clustered into four by using the
Durham-$k_T$ algorithm~\cite{Ref:KT}. 
The cluster is identified as $\gamma$, if the cluster contains only
$\gamma$'s.
The cluster is identified as $e^{\pm}$ or $\mu^{\pm}$, if the cluster
contains one and only one $e^{\pm}$ or $\mu^{\pm}$ and its $p_T^{}$ is
more than 95\% of the cluster. 
Otherwise, we identify the cluster as a jet. 
The jet is tagged as a $\tau$-jet, if the cluster contains one or
three charged particles and the sum of $p_T^{}$ of particles inside the
$R=0.15$ cone is more than 95\% of $p_T^{}$ of the cluster. 
The jet which has $B$-hadrons in the decay history of its constituent
particles is tagged as a $B$-jet with the probability of 65\%. 
The other jet which has $D$-hadrons in the decay history of its
constituent particles is tagged as a $B$-jet with the probability of
1\%. 
Other jets are tagged as $B$-jets with the probability of 0.1\%. 

For the $4b$ events, we take the events with four $B$-jets or three
$B$-jets plus one jet. 
With the above $B$-tagging probabilities, we find that the efficiency of
finding the $4b$ events is about 50\%. 
Absolute values of the 4-momenta of the four jets are rescaled so that
the sum of the 4-jet energy is equal to the collision energy and the sum of
the 3-momenta of the four jets vanishes. 
Then, the di-$B$-jet invariant mass $M_{BB}$ can be reconstructed, where
the pairs of the di-$B$-jet are chosen such that the difference of the
two invariant masses is minimal.
By fitting the distribution in a Breit-Wigner form, we get the width
$\Gamma_{\rm res}\simeq11$~GeV, which is assumed to be the systematical
resolution of the width measurement. 
We note that this value is roughly twice of that used in
Ref.~\cite{Ref:TanB}. 

For the $4\tau$ events, we take the events which contain four
$\tau$-jets or three $\tau$-jets with one charged lepton or two
$\tau$-jets with two same-sign charged leptons.
These signatures are expected to have small SM background
contributions. 
The efficiency of finding the $4\tau$ events is about 50\%.
Then, the 4-momenta of 4$\tau$'s can be reconstructed by rescaling the
absolute values of the 4-momenta of the four objects so that the energy
sum is equal to the collision energy and the sum of the 3-momenta
vanishes. 
The Di-$\tau$ invariant mass $M_{\tau\tau}$ is reconstructed, where
the pairs of di-$\tau$ are chosen such that the difference of the two
invariant masses is minimal with avoiding the same-sign charged leptons
to be paired.
By fitting the distribution in a Breit-Wigner form, we get
the width $\Gamma_{\rm res}\simeq7$~GeV. 
We note that the $\tau$-jet momentum is measured in a good accuracy by
the charged-tracks, while the accuracy of the collinear approximation in
the $\tau$ decays becomes a dominant source of the systematical
resolution of the width measurement.

%%%%%%%%%%%%%%%%%%%%%%%%%%%%%%%%%%%%%%%%%%%%%%%%%%%%%%%%%%%

%%%%%%%%%%%%%%%%%%%%%%%%%%%%%%%%%%%%%%%%%%%%%%%%%%%%%%%%%%%

\begin{thebibliography}{99}

\bibitem{Ref:atlas}
  ATLAS Collaboration,
  %``Observation of a new particle in the search for the Standard Model Higgs boson with the ATLAS detector at the LHC,''
  Phys.\ Lett.\ B {\bf 716}, 1 (2012). 
%  [arXiv:1207.7214 [hep-ex]].

\bibitem{Ref:cms}
  CMS Collaboration,
  %``Observation of a new boson at a mass of 125 GeV with the CMS experiment at the LHC,''
  Phys.\ Lett.\ B {\bf 716}, 30 (2012).
%  [arXiv:1207.7235 [hep-ex]].

\bibitem{Ref:atlas-comb}
 ATLAS Collaboration, 
  Report No. ATLAS-CONF-2013-034. 

\bibitem{Ref:cms-comb}
 CMS Collaboration, 
  Report No. CMS-PAS-HIG-12-045. 

\bibitem{Ref:HHG}
  J.~F.~Gunion, H.~E.~Haber, G.~Kane and S.~Dawson,
  The Higgs Hunter's Guide
  (Frontiers in Physics series, Addison-Wesley, Reading, MA, 1990).
  
\bibitem{Ref:GW}
  S.~L.~Glashow, S.~Weinberg,
  %``Natural Conservation Laws for Neutral Currents,''
  Phys.\ Rev.\  {\bf D15 } (1977)  1958.
  
\bibitem{Ref:2hdm}
  G.~C.~Branco, P.~M.~Ferreira, L.~Lavoura, M.~N.~Rebelo, M.~Sher and J.~P.~Silva,
  %``Theory and phenomenology of two-Higgs-doublet models,''
  Phys.\ Rept.\  {\bf 516}, 1 (2012).
%  [arXiv:1106.0034 [hep-ph]].

\bibitem{Ref:HK} 
  H.~E.~Haber and G.~L.~Kane,
  %``The Search for Supersymmetry: Probing Physics Beyond the Standard Model,''
  Phys.\ Rept.\  {\bf 117}, 75 (1985).

\bibitem{Ref:Djouadi2}
  A.~Djouadi,
  %``The Anatomy of electro-weak symmetry breaking. II. The Higgs bosons in the minimal supersymmetric model,''
  Phys.\ Rept.\  {\bf 459}, 1 (2008). 
%  [hep-ph/0503173].  

\bibitem{Ref:U1X}
  E.~Ma,
  %``New U(1) gauge symmetry of quarks and leptons,''
  Mod.\ Phys.\ Lett.\ A {\bf 17}, 535 (2002);
%  [hep-ph/0112232].

  E.~Ma and D.~P.~Roy,
  %``Heavy triplet leptons and new gauge boson,''
  Nucl.\ Phys.\ B {\bf 644}, 290 (2002).
%  [hep-ph/0206150].

\bibitem{Ref:AKS} 
  M.~Aoki, S.~Kanemura and O.~Seto,
  %``Neutrino mass, Dark Matter and Baryon Asymmetry via TeV-Scale Physics without Fine-Tuning,''
  Phys.\ Rev.\ Lett.\  {\bf 102}, 051805 (2009);
%  [arXiv:0807.0361 [hep-ph]].

  M.~Aoki, S.~Kanemura and O.~Seto,
  %``A Model of TeV Scale Physics for Neutrino Mass, Dark Matter and Baryon Asymmetry and its Phenomenology,''
  Phys.\ Rev.\ D {\bf 80}, 033007 (2009);
%  [arXiv:0904.3829 [hep-ph]].

  M.~Aoki, S.~Kanemura and K.~Yagyu,
  %``Triviality and vacuum stability bounds in the three-loop neutrino mass model,''
  Phys.\ Rev.\ D {\bf 83}, 075016 (2011).
%  [arXiv:1102.3412 [hep-ph]].

\bibitem{Ref:GHK} 
  H.~-S.~Goh, L.~J.~Hall and P.~Kumar,
  %``The Leptonic Higgs as a Messenger of Dark Matter,''
  JHEP {\bf 0905}, 097 (2009).
%  [arXiv:0902.0814 [hep-ph]].

\bibitem{Ref:BBES} 
  Y.~Bai, V.~Barger, L.~L.~Everett and G.~Shaughnessy,
  %``2HDM Portal Dark Matter: LHC data and the Fermi-LAT 135 GeV Line,''
  arXiv:1212.5604 [hep-ph].

\bibitem{Ref:CWWY}
  J.~Cao, P.~Wan, L.~Wu and J.~M.~Yang,
  %``Lepton-Specific Two-Higgs Doublet Model: Experimental Constraints and Implication on Higgs Phenomenology,''
  Phys.\ Rev.\ D {\bf 80}, 071701 (2009).
%  [arXiv:0909.5148 [hep-ph]].

\bibitem{Ref:mssm-HA-atlas}
  G.~Aad {\it et al.}  [ATLAS Collaboration],
  %``Search for the neutral Higgs bosons of the Minimal Supersymmetric Standard Model in $pp$ collisions at $\sqrt{s}=7$ TeV with the ATLAS detector,''
  JHEP {\bf 1302}, 095 (2013).
%  [arXiv:1211.6956 [hep-ex]].

\bibitem{Ref:mssm-HA-cms}
 CMS Collaboration, 
  Report No. CMS-PAS-HIG-12-050. 

\bibitem{Ref:2HDM-LHC} 
  Y.~Bai, V.~Barger, L.~L.~Everett and G.~Shaughnessy,
  %``The 2HDM-X and Large Hadron Collider Data,''
  arXiv:1210.4922 [hep-ph];

  A.~Celis, V.~Ilisie and A.~Pich,
  %``LHC constraints on two-Higgs doublet models,''
  arXiv:1302.4022 [hep-ph];

  C.~-W.~Chiang and K.~Yagyu,
  %``Implications of Higgs boson search data on the two-Higgs doublet models with a softly broken $Z_2$ symmetry,''
  arXiv:1303.0168 [hep-ph];
    
  P.~P.~Giardino, K.~Kannike, I.~Masina, M.~Raidal and A.~Strumia,
  %``The universal Higgs fit,''
  arXiv:1303.3570 [hep-ph];

  C.~-Y.~Chen, S.~Dawson and M.~Sher,
  %``Heavy Higgs Searches and Constraints on Two Higgs Doublet Models,''
  arXiv:1305.1624 [hep-ph];

  O.~Eberhardt, U.~Nierste and M.~Wiebusch,
  %``Status of the two-Higgs-doublet model of type II,''
  arXiv:1305.1649 [hep-ph];
 
  N.~Craig, J.~Galloway and S.~Thomas,
  %``Searching for Signs of the Second Higgs Doublet,''
  arXiv:1305.2424 [hep-ph].

\bibitem{Ref:bsg}
  M.~Ciuchini, E.~Franco, G.~Martinelli, L.~Reina, L.~Silvestrini,
  %``b ---> s gamma and b ---> s g: A Theoretical reappraisal,''
  Phys.\ Lett.\  {\bf B334 } (1994)  137-144;
%  [hep-ph/9406239].

  M.~Ciuchini, G.~Degrassi, P.~Gambino, G.~F.~Giudice,
  %``Next-to-leading QCD corrections to B ---> X(s) gamma: Standard model and two Higgs doublet model,''
  Nucl.\ Phys.\  {\bf B527 } (1998)  21-43;
%  [hep-ph/9710335].

  F.~Borzumati, C.~Greub,
  %``2HDMs predictions for anti-B ---> X(s) gamma in NLO QCD,''
  Phys.\ Rev.\  {\bf D58 } (1998)  074004;
%  [hep-ph/9802391]

  P.~Gambino, M.~Misiak,
  %``Quark mass effects in anti-B ---> X(s gamma),''
  Nucl.\ Phys.\  {\bf B611 } (2001)  338-366.
%  [hep-ph/0104034].

\bibitem{Ref:bsg2}
 M.~Misiak, H.~M.~Asatrian, K.~Bieri, M.~Czakon, A.~Czarnecki, T.~Ewerth, A.~Ferroglia, P.~Gambino {\it et al.},
  %``Estimate of B(anti-B ---> X(s) gamma) at O(alpha(s)**2),''
  Phys.\ Rev.\ Lett.\  {\bf 98 } (2007)  022002.
%  [hep-ph/0609232].

\bibitem{Ref:btaunu}
  W.~-S.~Hou,
  %``Enhanced charged Higgs boson effects in B- ---> tau anti-neutrino, mu anti-neutrino and b ---> tau anti-neutrino + X,''
  Phys.\ Rev.\  {\bf D48 } (1993)  2342-2344;

  Y.~Grossman, Z.~Ligeti,
  %``The Inclusive anti-B ---> tau anti-neutrino X decay in two Higgs doublet models,''
  Phys.\ Lett.\  {\bf B332 } (1994)  373-380;
%  [hep-ph/9403376, hep-ph/9403376].

  Y.~Grossman, H.~E.~Haber, Y.~Nir,
  %``QCD corrections to charged Higgs mediated b ---> c tau-neutrino decay,''
  Phys.\ Lett.\  {\bf B357 } (1995)  630-636.
%  [hep-ph/9507213].

\bibitem{Ref:tau}
  W.~Hollik, T.~Sack,
  %``Can a second Higgs doublet diminish the leptonic tau decay width?,''
  Phys.\ Lett.\  {\bf B284 } (1992)  427-430;

 M.~Krawczyk, D.~Temes,
  %``2HDM(II) radiative corrections in leptonic tau decays,''
  Eur.\ Phys.\ J.\  {\bf C44 } (2005)  435-446.
%  [hep-ph/0410248].

\bibitem{Ref:TypeX}
  V.~D.~Barger, J.~L.~Hewett, R.~J.~N.~Phillips,
  %``New Constraints On The Charged Higgs Sector In Two Higgs Doublet Models,''
  Phys.\ Rev.\  {\bf D41 } (1990)  3421;

  Y.~Grossman,
  %``Phenomenology of models with more than two Higgs doublets,''
  Nucl.\ Phys.\  {\bf B426 } (1994)  355-384.
%  [hep-ph/9401311].

\bibitem{Ref:AKTY}
  M.~Aoki, S.~Kanemura, K.~Tsumura, K.~Yagyu,
  %``Models of Yukawa interaction in the two Higgs doublet model, and their collider phenomenology,''
  Phys.\ Rev.\  {\bf D80 } (2009)  015017.
%  [arXiv:0902.4665 [hep-ph]]. 
  
\bibitem{Ref:Su}
  S.~Su and B.~Thomas,
  %``The LHC Discovery Potential of a Leptophilic Higgs,''
  Phys.\ Rev.\ D {\bf 79}, 095014 (2009).
%  [arXiv:0903.0667 [hep-ph]].
  
\bibitem{Ref:Logan}
 H.~E.~Logan and D.~MacLennan,
  %``Charged Higgs phenomenology in the lepton-specific two Higgs doublet model,''
  Phys.\ Rev.\ D {\bf 79}, 115022 (2009).
%  [arXiv:0903.2246 [hep-ph]].

\bibitem{Ref:h-BR}
  H.~Ono and A.~Miyamoto,
  %``A study of measurement precision of the Higgs boson branching ratios at the International Linear Collider,''
  Eur.\ Phys.\ J.\ C {\bf 73}, 2343 (2013). 

\bibitem{Ref:ilc-TDR}
  J.~Brau, P.~Grannis, M.~Harrison, M.~Peskin, M.~Ross and H.~Weerts,
  %``The International Linear Collider,''
  arXiv:1304.2586 [physics.acc-ph].

\bibitem{Ref:ilc-Peskin}
  M.~E.~Peskin,
  %``Comparison of LHC and ILC Capabilities for Higgs Boson Coupling Measurements,''
  arXiv:1207.2516 [hep-ph].
  
\bibitem{Ref:TanB}
  V.~D.~Barger, T.~Han and J.~Jiang,
  %``Tan beta determination from heavy Higgs boson production at linear colliders,''
  Phys.\ Rev.\ D {\bf 63}, 075002 (2001); 
%  [hep-ph/0006223].

  J.~F.~Gunion, T.~Han, J.~Jiang and A.~Sopczak,
  %``Determining tan beta with neutral and charged Higgs bosons at a future e + e- linear collider,''
  Phys.\ Lett.\ B {\bf 565}, 42 (2003).
%  [hep-ph/0212151].

\bibitem{Ref:mssm-H+}
  G.~Aad {\it et al.}  [ATLAS Collaboration],
  %``Search for charged Higgs bosons decaying via $H^{+} \to \tau \nu$ in top quark pair events using $pp$ collision data at $\sqrt{s}=7$ TeV with the ATLAS detector,''
  JHEP {\bf 1206}, 039 (2012);
%  [arXiv:1204.2760 [hep-ex]].
%
%  G.~Aad {\it et al.}  [ATLAS Collaboration],
  %``Search for a light charged Higgs boson in the decay channel H+->csbar in ttbar events using pp collisions at sqrt(s) = 7 TeV with the ATLAS detector,''
  arXiv:1302.3694 [hep-ex].
  
\bibitem{Ref:KOSY} 
  S.~Kanemura, Y.~Okada, E.~Senaha and C.~-P.~Yuan,
  %``Higgs coupling constants as a probe of new physics,''
  Phys.\ Rev.\ D {\bf 70}, 115002 (2004).
%  [hep-ph/0408364].

\bibitem{Ref:Uni-2hdm}
  S.~Kanemura, T.~Kubota and E.~Takasugi,
  %``Lee-Quigg-Thacker bounds for Higgs boson masses in a two doublet model,''
  Phys.\ Lett.\ B {\bf 313}, 155 (1993); 
%  [hep-ph/9303263].

  A.~G.~Akeroyd, A.~Arhrib and E.~-M.~Naimi,
  %``Note on tree level unitarity in the general two Higgs doublet model,''
  Phys.\ Lett.\ B {\bf 490}, 119 (2000).
%  [hep-ph/0006035].

\bibitem{Ref:rho-2hdm}
  D.~Toussaint,
  %``Renormalization Effects From Superheavy Higgs Particles,''
  Phys.\ Rev.\ D {\bf 18}, 1626 (1978);

\bibitem{Ref:rho2-2hdm}
  S.~Bertolini,
  %``Quantum Effects In A Two Higgs Doublet Model Of The Electroweak Interactions,''
  Nucl.\ Phys.\ B {\bf 272}, 77 (1986).  

\bibitem{Ref:rho3-2hdm}
  W.~Hollik,
  %``NONSTANDARD HIGGS BOSONS IN SU(2) x U(1) RADIATIVE CORRECTIONS,''
  Z.\ Phys.\ C {\bf 32}, 291 (1986);

%  W.~Hollik,
  %``Radiative Corrections With Two Higgs Doublets At Lep / Slc And Hera,''
  Z.\ Phys.\ C {\bf 37}, 569 (1988).
  
\bibitem{Ref:KOTT} 
  S.~Kanemura, Y.~Okada, H.~Taniguchi and K.~Tsumura,
  %``Indirect bounds on heavy scalar masses of the two-Higgs-doublet model in light of recent Higgs boson searches,''
  Phys.\ Lett.\ B {\bf 704}, 303 (2011).
%  [arXiv:1108.3297 [hep-ph]].

\bibitem{Ref:Zh-lep}
ILD Concept Group, The International Large Detector: Letter of Intent, KEK Report 2009-6.

\bibitem{Ref:Zh-had}
A. Yamamoto, presentation at the Asian Physics and Software Meeting, June, 2012.
	
\bibitem{Ref:KTY}
  S.~Kanemura, K.~Tsumura and H.~Yokoya,
  %``Multi-tau-lepton signatures at the LHC in the two Higgs doublet model,''
  Phys.\ Rev.\ D {\bf 85}, 095001 (2012); 
%  [arXiv:1111.6089 [hep-ph]].

%  S.~Kanemura, K.~Tsumura and H.~Yokoya,
  %``Multi-Tau Lepton Signatures in Leptophilic Two Higgs Doublet Model at the ILC,''
  Proceedings for LCWS11, 26-30 Sep 2011. Granada, Spain, 
  arXiv:1201.6489 [hep-ph].

\bibitem{Ref:PYTHIA}
  T.~Sjostrand, S.~Mrenna and P.~Z.~Skands,
  %``PYTHIA 6.4 Physics and Manual,''
  JHEP {\bf 0605} (2006) 026.
%  [hep-ph/0603175].

\bibitem{Ref:FastJet}
  M.~Cacciari, G.~P.~Salam and G.~Soyez,
  %``FastJet User Manual,''
  Eur.\ Phys.\ J.\ C {\bf 72} (2012) 1896.
%  [arXiv:1111.6097 [hep-ph]].
  
  
  
  
  
  
  
  
  
  
  
  
  
  

\bibitem{Ref:KT}
  S.~Catani, Y.~L.~Dokshitzer, M.~Olsson, G.~Turnock and B.~R.~Webber,
  %``New clustering algorithm for multi - jet cross-sections in e+ e- annihilation,''
  Phys.\ Lett.\ B {\bf 269} (1991) 432.













\end{thebibliography}
\end{document}